\documentclass[runningheads]{llncs}
\usepackage{graphicx}
\usepackage{multirow}
\usepackage{eucal}
\usepackage{multicol}
\usepackage{booktabs}
\usepackage{subfig}
\usepackage{amsmath}
\usepackage{gensymb}
\usepackage{amssymb}
\usepackage{color}
\newcommand{\tabincell}[2]{\begin{tabular}{@{}#1@{}}#2\end{tabular}}  
\begin{document}
\title{Generalisable Cardiac Structure Segmentation via Attentional and Stacked Image Adaptation}
\titlerunning{Generalisable Cardiac Structure Segmentation via Attentional and Stacked Image Adaptation}
\author{Hongwei Li\inst{1} \and
   Jianguo Zhang\inst{2} \and
   Bjoern Menze\inst{1} }
 \authorrunning{H. Li et al.}
 \institute{Department of Computer Science, Technical University of Munich, Germany \and  Department of Computer Science and Engineering, 
 \\ Southern University of Science and Technology, Shenzhen, China
\\ Emails: \email{\{hongwei.li, bjoern.menze\}@tum.de}; \email{zhangjg@sustech.edu.cn}}
\maketitle              
\begin{abstract}
Tackling domain shifts in multi-centre and multi-vendor data sets remains challenging for cardiac image segmentation. In this paper, we propose a generalisable segmentation framework for cardiac image segmentation in which multi-centre, multi-vendor, multi-disease datasets are involved. A generative adversarial networks with an attention loss was proposed to translate the images from existing source domains to a target domain, thus to generate good-quality synthetic cardiac structure and enlarge the training set. A stack of data augmentation techniques was further used to simulate real-world transformation to boost the segmentation performance for unseen domains. We achieved an average Dice score of \textbf{90.3\%} for the left ventricle, \textbf{85.9\%} for the myocardium, and \textbf{86.5\%} for the right ventricle on the hidden validation set across four vendors.  We show that the domain shifts in heterogeneous cardiac imaging datasets can be drastically reduced by two aspects: 1) good-quality synthetic data by learning the underlying target domain distribution, and 2) stacked classical image processing techniques for data augmentation. 

\end{abstract}

\section{Introduction}
Fully automatic cardiac segmentation methods can help clinicians to quantify the heart structure (e.g. left ventricle (LV), myocardium (MYO), and right ventricle (RV)) from cardiac magnetic resonance (CMR) images for diagnosis of multiple heart diseases \cite{bernard2018deep}. 
Deep learning-based methods have shown promising avenues for cardiac image segmentation \cite{chen2020deep}. However,
existing work \cite{yan2020mri} have shown that the segmentation performance of such methods may drop in when they are directly tested to scans acquired from different centres or vendors. The degradation of performance is not only caused by the varying cardiac morphology but also the differences of acquisition parameter, resolution, intensity distribution, etc. \cite{petitjean2011review} as shown in Fig. \ref{samples}. All these factors pose obstacles for deploying deep learning-based segmentation algorithms in real-world clinical practice. 

\begin{figure}
	\centering
    \includegraphics[height=0.37\textwidth,width=\textwidth]{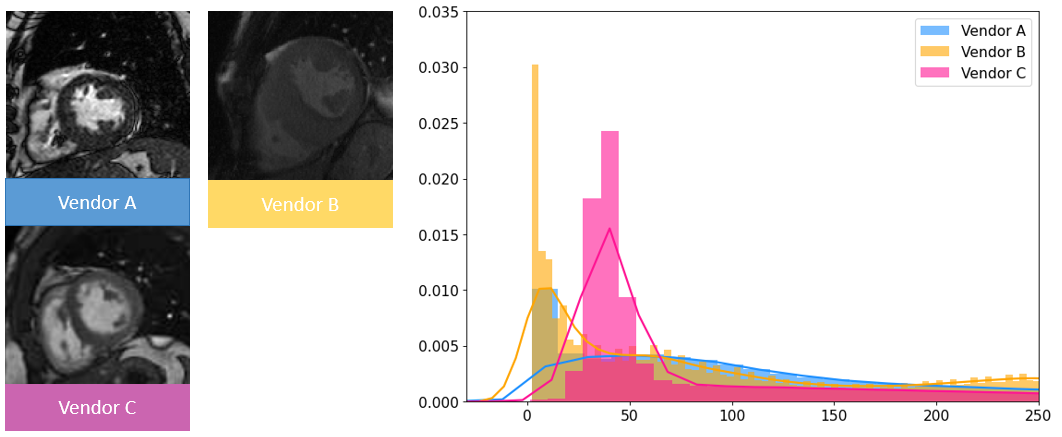}
    \caption{The differences of image appearance, resolution and intensity distributions in the cardiac structures (i.e. LV, MYO, RV) of three vendors.}
	\label{samples}
\end{figure}

One simple way to address the above issues
is to fine-tune a model learned from some datasets (source domains) with extra labelled data from another dataset (target domain). However, collecting sufficient pixel-level labelled medical data for all centres and vendors is extremely difficult which require fully clinical studies. To mitigate these issues, domain adaptation methods have been proposed to generalise one algorithm trained on some datasets (source domain) with additional data (either labelled or not labelled) from another dataset
(target domain) \cite{chen2019synergistic,zhuang2020cardiac,yan2019domain}. Data augmentation-based methods are further proposed to enhance the generalisability of cardiac image segmentation models \cite{zhang2020generalizing,chen2019improving}. Generally, it is clinically relevant to explore how to learn a generalisable model that can be successfully applied to other datasets without additional model tuning. 

In this work, we present a fully automatic segmentation framework to segment three cardiac structures (i.e. LV, MYO, and RV) and to mitigate the above issues caused by domain shifts. It is achieved by leveraging generative adversarial networks to transfer image style and stacked image processing techniques to augment the training samples and thus to generalise the segmentation model to unseen domains. Specifically, our approach mainly consists of three modules:\\
\noindent{(i) A target domain transfer network.} This module is used for learning the underlying intensity distribution of the unlabeled vendor and translate the labeled vendor to the target vendor, to augment the training set with \textbf{synthetic vendor-C-like} images and annotation from vendor \emph{A} and \emph{B }. Specifically we develop an attention-GANs with a focus on the cardiac structure. \\
\noindent{(ii) A stacked image transformation function. To simulate real-world testing conditions and to increase data variations, we apply a stack of six spatial and intensity transformations to overcome the domain shifts.} \\
\noindent{(iii) A segmentation model. A residual U-shape convolutional neural network with dilated convolutions \cite{li2018automatic} with a larger receptive field compared with U-Net is used to perform segmentation. It can capture richer context information with less parameters and is trained with all the data simulated above.}

\section{Methodology}

\begin{figure}[t]
	\centering
    \includegraphics[height=0.36\textwidth,width=\textwidth]{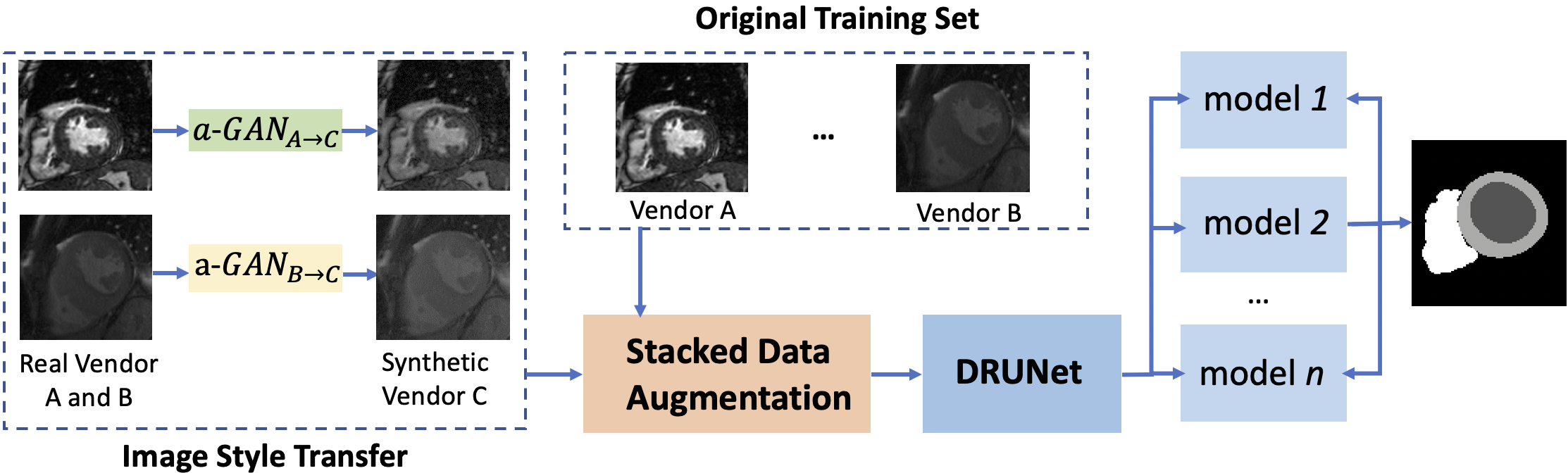}
    \caption{Schematic view of our approach. It consists of three modules: a) image style transfer involving an attention GAN with a focus on cardiac structure, b) stacked data augmentation of several image transformations, and c) a segmentation network called \emph{DRUNET} with lighter number of weight but capture rich context information. }
	\label{framework}
\end{figure}

Given the challenge setting \footnote{https://www.ub.edu/mnms/}, the proposed method aims at learning an generalisable segmentation model for labelled vendor \emph{A} and labelled vendor \emph{B}:\{(x$_{A}$, y$_{A}$), (x$_{B}$, y$_{B}$)\}, unlabelled vendor \emph{C}: \{x$_{C}$\} where 25 scans were given but without labels, and the unseen vendor \emph{D}, where the data was hidden. 

As shown in Fig. \ref{framework}, our framework mainly included three modules as mentioned above. Specifically, the image style transfer module translated the images from vendor \emph{A} and vendor \emph{B} to unlabeled vendor \emph{C} and further augmented the training set with {synthetic vendor-C-like} images and the annotations from vendor \emph{A} and \emph{B}. We enhanced the image quality by introducing an attention loss. The stacked data augmentation module aimed to further increase the data variation by employing several intensity and spatial transformation on the original and synthetic data after the first module. Last, the segmentation models were trained with all the original and synthetic data and performed inference with an ensemble model.

\subsubsection{Generative Adversarial Network with Attention.}
\begin{figure}
	\centering
    \includegraphics[height=0.35\textwidth,width=0.9\textwidth]{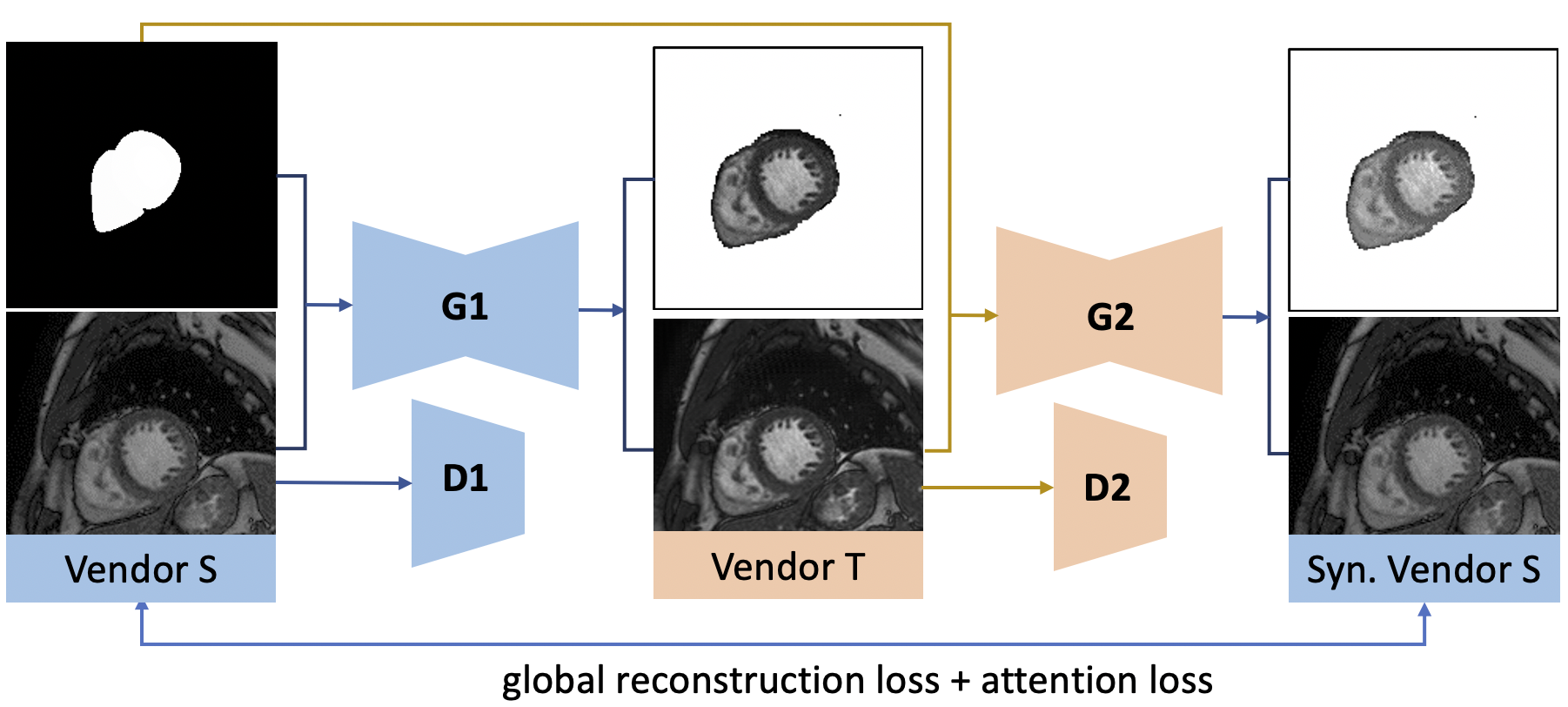}
    \caption{Overview of the image translation network. An attention- reconstruction loss was introduced to enhance the image quality of cardiac region. }
	\label{GANs}
\end{figure}

We adopted the CycleGAN \cite{zhu2017unpaired} as the basic architecture for one-to-one image translation. This included a pair of generators $\{G_{1}, G_{2}\}$ and a pair of domain discriminators $\{D_{1}, D_{2}\}$. As illustrated in Figure \ref{GANs}, generator $G_{1}$ aimed to translate images from a source vendor to a target vendor while generator $G_{2}$ aimed to translate from the target one to the source one. Discriminators $\{D_{1}, D_{2}\}$ are trained to distinguish if the generated images are real or fake in the two domains respectively. In a min-max game, the generators try to fool the discriminators by good image translation. However, since the cardiac structures were of interests, we further introduced a attention-reconstruction loss with a focus on cardiac structure:
\begin{equation}
	\mathcal{L}_{a\_rec} = \mathbb{E}_{x \sim \mathcal{S}}[||(x -G_{2}(G_{1}(x)))\odot m||_{1}] + \mathbb{E}_{y \sim \mathcal{T}}[||y-G_{1}(G_{2}(y))||_{1}]
	\end{equation}
Notably the mask \emph{m} was only applied to the source domains where the labels are available.
Thus, in addition to the original global reconstruction loss $\mathcal{L}_{g\_rec}$, we used a composite loss function which consists of two loss terms: $\mathcal{L}_{rec} =  \mathcal{L}_{g\_rec} + \lambda \mathcal{L}_{a\_rec}$. In
our experiments, we set $\lambda$ = 0.5 to balance the contribution of the two losses.
\vspace{-0.1cm}
\subsubsection{Stacked Augmentation.}
We proposed a sequence of $k$ stacked image transformations $f(\cdot)$ to simulate image distributions for unseen domains. 
Given training data $x_{t}$ and the corresponding label $y_{t}$, augmented data $\hat{x}_{S}$ and the paired label $\hat{y}_{t}$ can be generated after $k$ transformations.
 \begin{equation}
 	(\hat{x}_{t}, \hat{y}_{t}) = f^{k}(f^{k-1}(...f^1(x_{t}, y_{t})))
 \end{equation}
We selected three intensity transformations and three spatial transformations: \\
\noindent{(1) image blurring and sharpening. Gaussian filtering was utilised to smooth the image to simulate blurry produced by motions. The standard deviation of a Gaussian kernel is ranged between [0.1, 2]. Meanwhile, we sharpened the image by using a unsharp masking. \\ 
\noindent{(2) intensity perturbation. The intensity range was shifted with a magnitude range between [-0.05, 0.05]}.\\
\noindent{(3) gamma correction.} This was used to adjust the contrast of the image with a range between [0.6, 1.7].\\
\noindent{(4) shearing.} This was aimed to equip the network with variance to deformations with a magnitude range between [-0.1, 0.1] was used for both images and masks.\\
\noindent{(5) rotation. A range between [-15\degree, 15\degree ] was used for both images and masks.} \\
\noindent{(6) scaling. This enforces the network to be scale-invariant to resolution. A magnitude range between [-0.1, 0.1]} was used for both images and masks. \\
\subsubsection{Segmentation Network.}
We adopted a top-perform 2D architecture named Dilated Residual U-Net (DRUNet) \cite{li2018automatic}, which was used for both brain and cardiac segmentation. 
DRUNet exploited the inherent advantages of the skip connections, residual learning and dilated convolutions to capture rich context information with a minimal number of trainable parameters. The network was trained with a weight cross entropy loss function.

\section{Experiments}
\begin{table*}[htpb]
  \caption[table: Data Properties]{Multi-vendor datasets. Resolutions of scans from the same vendor are even different.}
  \label{tab:datasets}
  \centering
   \begin{tabular}{l c c c c }
    \hline 
     Vendor& A & B & C  & D \\
    \hline
      {\tabincell{l}{Numbers\\(training/test)}}   & 75/50& 75/50  & 25/50 &0/50\\
     \hline
      {\tabincell{l}{Annotation\\Availability~(\%)}}  & yes& yes  & no & no\\
     \hline
  \end{tabular}
\end{table*}


\subsection{Experimental Setting}

\noindent\textbf{Datasets.} The released training set consists of 150 annotated scans from two different MRI vendors (75 for each) and 25 unannotated scans from a third vendor as shown in Table \ref{tab:datasets}. The CMR scans have been segmented by experienced clinicians, with contours for the left (LV) and right ventricle (RV) blood pools, and the left ventricular myocardium (MYO). The segmentation pipeline was evaluated on it and the results on the hidden validation set provided by the challenge organisers were presented. To optimise the segmentation model We use four labeled scans (2 from vendor A and 2 from vendor B) as a validation set, and the remains as a training set. For the final submission, we used all the whole released training set. 

\noindent\textbf{Pre- and post- processing.}  The pre-processing of all the images (including the hidden test set) were performed in a slice-wise manner by three steps. First, non-local means denoising was performed for each slice to reduce the noise level considering that the image quality from multiple centres are diverse; second, the intensity range was normalised to [0,~1] to facilitate the model training; third, the images and masks are cropped or padded to [256,~256]. For post-processing, we performed connected component analysis and removed small structures with less than 30 voxels. 

\noindent\textbf{Network training.}  (1) For the image translation network, we used the CycleGAN implementation for the one-to-one mappings: \emph{A} $\rightarrow$ \emph{C}, and  \emph{B} $\rightarrow$ \emph{C}. Network configuration and hyper-parameters were kept the same as in \cite{zhu2017unpaired} except the input and output images are single-channel 2D images. It was trained for 100 epochs with a batch size of 5 involving around 3500 images for each vendor including scans from multiple time points. (2) For training the segmentation model, the weights for background, LV, MYO, and
RV in the weighted cross entropy loss were empirically set to 0.19 : 0.24 : 0.31 : 0.26 based on the performance on the validation set. The algorithm was implemented using python and Tensorflow and was trained for
100 epochs in total on an NVIDIA® Titan V GPU. The training of the segmentation model took around 5 hours. 

\subsection{Results}
We conducted three experiments to illustrate the effectiveness of our approach. First, we trained the DRUNet on solely the labeled datasets: vendor \emph{A} and \emph{B}, referred as \emph{baseline} in Table \ref{tab:results_Dice}. Second, we used the attention-GANs to generate good-quality vendor \emph{C}-like images and include those synthetic images and their corresponding labels for training, referred as \emph{baseline+a-GANs} in Table \ref{tab:results_Dice}. Lastly, we further incorporated the stacked image transformation and train the model from scratch, referred as \emph{ours} in the table. We found that after including the stacked image transformation, we drastically improved the segmentation performance on the hidden vendor \emph{D}, e.g. Dice for RV is improved from 14.5\% to 72.7\%. On the unlabeled vendor \emph{C}, we achieved average Dice score of 86.8\% for the left ventricle, 83.4\% for the myocardium, and
83.2\% for the right ventricle; on the hidden vendor, we achieved average Dice score of \textbf{89.3\%} for the left ventricle, \textbf{83.4\%} for the myocardium, and
72.7\% for the right ventricle.  Qualitative segmentation result from vendor \emph{C} is shown in Fig. \ref{samples}.

\begin{table}
\centering
\caption{Average Dice scores of all vendors, the highest performance in each class is highlighted.}
\begin{tabular}{|c|c|c|c|c|c|c|c|c|c|c|c|c|c|c|c|}
\hline
\multirow{2}{*}{Method}      &  \multicolumn{3}{c|}{Dice$_A$(\%)} &\multicolumn{3}{c|}{Dice$_B$(\%)} & \multicolumn{3}{c|}{Dice$_C$(\%)} &\multicolumn{3}{c| }{Dice$_D$(\%)}  \\
\cline{2-13}
 ~ &LV & MYO&RV& LV& MYO & RV & LV & MYO & RV  &LV & MYO&RV\\
\hline
Baseline &85.7 &77.1 & 66.6 & 92.2 &  83.9& 87.7  & 86.0  & 81.0 & 76.5 & 72.3  & 51.7 &14.5 \\

Baseline+a-GANs & 88.5  & 81.6 &71.8  & \textbf{94.2} & 86.6 & \textbf{91.5} & \textbf{87.7}  & \textbf{84.5} & 80.1 & 65.9 & 58.0 &13.3 \\
Ours (a-GANs+Stacked) &  \textbf{90.5} & \textbf{84.1} & \textbf{85.1}  & {93.6} & \textbf{87.5} & {91.1} & {86.8}  & {83.4} & \textbf{83.2} & \textbf{89.3} & \textbf{81.4} & \textbf{72.7}  \\
\hline
\end{tabular}
\label{tab:results_Dice}
\end{table}
\vspace{-0.2cm}
\begin{table}
\centering
\caption{Average Hausdorff distance (HD) of all vendors, the highest performance in each class is highlighted.}
\begin{tabular}{|c|c|c|c|c|c|c|c|c|c|c|c|c|c|c|c|}
\hline
\multirow{2}{*}{Method}      &  \multicolumn{3}{c|}{HD$_A$(mm)} &\multicolumn{3}{c|}{HD$_B$(mm)} & \multicolumn{3}{c|}{HD$_C$(mm)} &\multicolumn{3}{c| }{HD$_D$(\%)}  \\
\cline{2-13}
 ~ &LV & MYO&RV& LV& MYO & RV & LV) & MYO & RV  &LV & MYO&RV\\
\hline
Baseline &23.7  & 37.0 & 44.11 & 14.0 & 20.7 &23.4 &17.7 &19.4 &31.5 &27.5 &35.8  &61.8 \\

Baseline+a-GANs & 21.4  &31.2  &21.1  &\textbf{7.5}  &{11.7}  &12.4  &14.4  &16.3  &\textbf{16.7 } &22.3  &30.3  &42.0 \\
Ours (a-GANs+Stacked) &  \textbf{15.8} & \textbf{16.7} & \textbf{16.2}  & {7.9} & \textbf{11.0} & \textbf{11.5} & \textbf{10.9}  & \textbf{15.0} & {23.6} & \textbf{17.3} & \textbf{24.6} & \textbf{17.6}  \\

\hline
\end{tabular}
\label{tab:results_Dice}
\end{table}
\vspace{-0.2cm}

\begin{figure}
	\centering
    \includegraphics[height=0.75\textwidth,width=\textwidth]{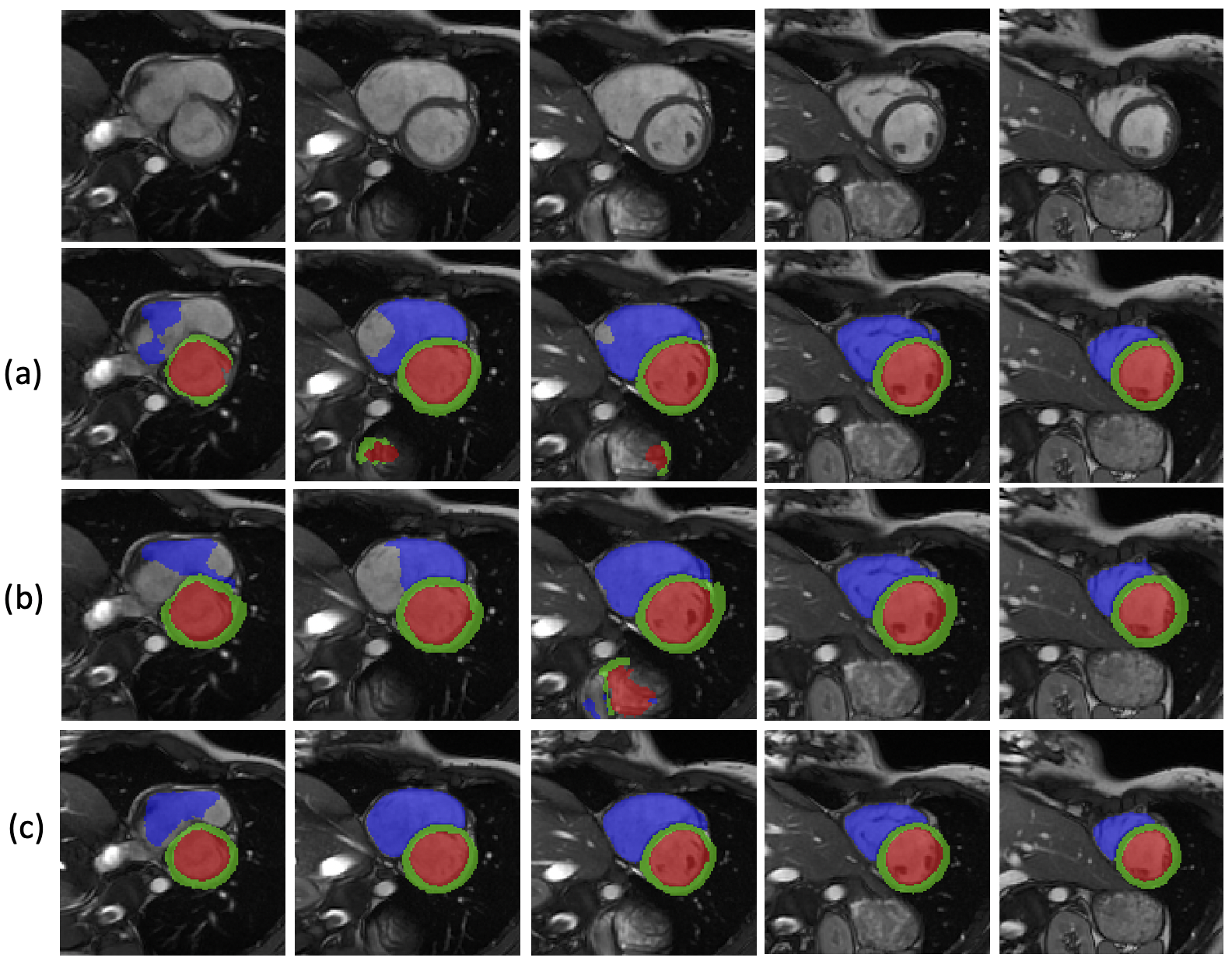}
    \caption{Qualitative segmentation result of one subject from vendor \emph{C}. a) baseline results; b) results of baseline+a-GANs; c) results of our method. Please note that the ground truth for vendor \emph{C} is not available.}
	\label{samples}
\end{figure}

\section{Conclusion}
We proposed a cardiac structure segmentation framework and segmented three structures: LV, MYO, and RV. 
We demonstrated that domain shifts in heterogeneous cardiac imaging datasets can be drastically reduced by two aspects: 1) good-quality synthetic data by learning the underlying target domain distribution, and 2) stacked classical image transformation techniques for data augmentation. 
\bibliographystyle{splncs04}
\bibliography{egbib}

\begin{thebibliography}{10}
\providecommand{\url}[1]{\texttt{#1}}
\providecommand{\urlprefix}{URL }
\providecommand{\doi}[1]{https://doi.org/#1}

\bibitem{bernard2018deep}
Bernard, O., Lalande, A., Zotti, C., Cervenansky, F., Yang, X., Heng, P.A.,
  Cetin, I., Lekadir, K., Camara, O., Ballester, M.A.G., et~al.: Deep learning
  techniques for automatic mri cardiac multi-structures segmentation and
  diagnosis: is the problem solved? IEEE transactions on medical imaging
  \textbf{37}(11),  2514--2525 (2018)

\bibitem{chen2019improving}
Chen, C., Bai, W., Davies, R.H., Bhuva, A.N., Manisty, C., Moon, J.C., Aung,
  N., Lee, A.M., Sanghvi, M.M., Fung, K., et~al.: Improving the
  generalizability of convolutional neural network-based segmentation on cmr
  images. arXiv preprint arXiv:1907.01268  (2019)

\bibitem{chen2020deep}
Chen, C., Qin, C., Qiu, H., Tarroni, G., Duan, J., Bai, W., Rueckert, D.: Deep
  learning for cardiac image segmentation: A review. Frontiers in
  Cardiovascular Medicine  \textbf{7}, ~25 (2020)

\bibitem{chen2019synergistic}
Chen, C., Dou, Q., Chen, H., Qin, J., Heng, P.A.: Synergistic image and feature
  adaptation: Towards cross-modality domain adaptation for medical image
  segmentation. In: Proceedings of the AAAI Conference on Artificial
  Intelligence. vol.~33, pp. 865--872 (2019)

\bibitem{li2018automatic}
Li, H., Zhygallo, A., Menze, B.: Automatic brain structures segmentation using
  deep residual dilated u-net. In: International MICCAI Brainlesion Workshop.
  pp. 385--393. Springer (2018)

\bibitem{petitjean2011review}
Petitjean, C., Dacher, J.N.: A review of segmentation methods in short axis
  cardiac mr images. Medical image analysis  \textbf{15}(2),  169--184 (2011)

\bibitem{yan2020mri}
Yan, W., Huang, L., Xia, L., Gu, S., Yan, F., Wang, Y., Tao, Q.: Mri
  manufacturer shift and adaptation: Increasing the generalizability of deep
  learning segmentation for mr images acquired with different scanners.
  Radiology: Artificial Intelligence  \textbf{2}(4),  e190195 (2020)

\bibitem{yan2019domain}
Yan, W., Wang, Y., Gu, S., Huang, L., Yan, F., Xia, L., Tao, Q.: The domain
  shift problem of medical image segmentation and vendor-adaptation by
  unet-gan. In: International Conference on Medical Image Computing and
  Computer-Assisted Intervention. pp. 623--631. Springer (2019)

\bibitem{zhang2020generalizing}
Zhang, L., Wang, X., Yang, D., Sanford, T., Harmon, S., Turkbey, B., Wood,
  B.J., Roth, H., Myronenko, A., Xu, D., et~al.: Generalizing deep learning for
  medical image segmentation to unseen domains via deep stacked transformation.
  IEEE Transactions on Medical Imaging  (2020)

\bibitem{zhu2017unpaired}
Zhu, J.Y., Park, T., Isola, P., Efros, A.A.: Unpaired image-to-image
  translation using cycle-consistent adversarial networks. In: Proceedings of
  the IEEE international conference on computer vision. pp. 2223--2232 (2017)

\bibitem{zhuang2020cardiac}
Zhuang, X., Xu, J., Luo, X., Chen, C., Ouyang, C., Rueckert, D., Campello,
  V.M., Lekadir, K., Vesal, S., RaviKumar, N., et~al.: Cardiac segmentation on
  late gadolinium enhancement mri: A benchmark study from multi-sequence
  cardiac mr segmentation challenge. arXiv preprint arXiv:2006.12434  (2020)

\end{thebibliography}
\end{document}